\documentclass[prl,nofootinbib,superscriptaddress,twocolumn]{revtex4}
\usepackage[a4paper,left=1.5cm,right=1.5cm,top=3cm,bottom=3cm]{geometry}

\usepackage{amssymb,amsmath,amsfonts}
\usepackage[dvipsnames]{xcolor}
\usepackage{graphicx}
\usepackage{longtable}
\usepackage{verbatim}
\usepackage{color}
\usepackage{hyperref}
\hypersetup{colorlinks, citecolor=MidnightBlue, linkcolor=bluscuro, urlcolor=Cyan}
\definecolor{rossos}{cmyk}{0,1,1,0.55}
\definecolor{bluscuro}{rgb}{0.15, 0.2, .85}
\definecolor{bluchiaro}{cmyk}{1,.3,0.,0.1}

\graphicspath{{./Figures/}}

\newcommand{\be}{\begin{equation}}
\newcommand{\ee}{\end{equation}}
\newcommand{\bea}{\begin{eqnarray}}
\newcommand{\eea}{\end{eqnarray}}
\newcommand{\beq}{\begin{equation}}
\newcommand{\eeq}{\end{equation}}
\def\beqa{\begin{eqnarray}}

\def\eeqa{\end{eqnarray}}

\def\lsim{\mathrel{\rlap{\lower4pt\hbox{\hskip0.5pt$\sim$}}
    \raise1pt\hbox{$<$}}}         
\def\gsim{\mathrel{\rlap{\lower4pt\hbox{\hskip0.5pt$\sim$}}
    \raise1pt\hbox{$>$}}}         
\newcommand{\TRH}{T_\textrm{RH}}
\newcommand{\hc}{h_\textrm{c}}
\newcommand{\te}{t_\textrm{e}}
\renewcommand{\S}{S}
\newcommand{\mS}{m_\S}

\newcommand{\Nstar}{\overline{N}_*}
\newcommand{\hstar}{\overline{h}_*}
\newcommand{\Pz}{{\cal P}_\zeta}

\newcommand{\arXiv}[2]{\href{http://arxiv.org/pdf/#1}{{\tt [#2/#1]}}}
\newcommand{\arXivold}[1]{\href{http://arxiv.org/pdf/#1}{{\tt [#1]}}}

\begin{document}

\title{Primordial Black Holes from  Higgs
Vacuum Instability:\\
Avoiding Fine-tuning through an Ultraviolet Safe Mechanism}
\author{J.~R.~Espinosa }
\address{Institut de F\'{\i}sica d'Altes Energies (IFAE), The Barcelona Institute of Science and Technology (BIST),
Campus UAB, 08193 Bellaterra, Barcelona, Spain}
\address{ICREA, Instituci\'o Catalana de Recerca i Estudis Avan\c{c}ats, \\ 
Passeig de Llu\'{\i}s Companys 23, 08010 Barcelona, Spain}
\author{D.~Racco }
\author{A.~Riotto}
\address{D\'epartement de Physique Th\'eorique and Centre for Astroparticle Physics (CAP), Universit\'e de Gen\`eve, 24 quai E. Ansermet, CH-1211 Geneva, Switzerland}

\date{\today}

\begin{abstract}
\noindent
We have recently proposed the idea that dark matter in our universe is formed  by primordial black holes generated by   Standard Model  Higgs fluctuations
during inflation  and thanks to the fact that the Standard Model Higgs potential develops an instability at a scale of the order of $10^{11}$ GeV. In this sense, dark matter does not need any physics beyond the Standard Model, although
 the mechanism needs fine-tuning to avoid the overshooting of the Higgs into the dangerous AdS vacuum.
We show how such fine-tuning can be naturally  avoided by coupling the Higgs to a very heavy scalar with mass $\gg 10^{11}$ GeV that 
stabilises the potential in the deep ultraviolet,  but preserving  the basic feature of the mechanism which is built within the Standard Model.

\end{abstract}

\maketitle

\paragraph{Introduction.}
In a recent paper \cite{us}  we have proposed the idea that the Standard Model  (SM) of weak interactions might provide a candidate for dark matter, without the need of physics beyond the SM, under the form of Primordial Black Holes (PBHs).
The scenario is the following. As is well-known, the Higgs potential develops an instability at large values of the field  \cite{instab2,instab}. For instance, the Higgs potential becomes negative for Higgs field values of the order of $10^{11}$ GeV if we select the present central values of the Higgs and top masses. At this scale, the quartic coupling $\lambda$ in the Higgs potential turns negative.
During inflation, which is necessary to explain the anisotropies in the cosmic microwave background radiation as well as to provide the seeds for the large-scale structure, the Higgs field is subject to quantum fluctuations as any other field lighter than the Hubble rate $H$ \cite{espinosa}. The Higgs field performs a random walk which allows it to surmount the barrier of its effective potential. 

After surmounting the barrier, the Higgs continues its random walk till it reaches the field value at which 
the classical motion dominates over the quantum jumps. This happens when $\Delta_c h\simeq -V'(h)/3H\gsim (H/2\pi)$, or 
\be
h^3\gsim \frac{3H^3}{2\pi\lambda},
\ee
where for simplicity we have approximated the Higgs potential as $V(h)=-\lambda h^4/4$. The Higgs field starts a slow-roll phase till it reaches the value 
$h^2\sim 3H^2/\lambda$ at which slow- roll ends and the Higgs starts moving rapidly towards the bottom of the potential.
The would-be inevitable destiny of the Higgs field to create a crunching anti de Sitter region is however changed by the fact that inflation ends, the universe reheats at a temperature $\TRH$ and the plasma particles generated during reheating couple to the Higgs,  providing to its effective potential a large positive mass that can stop the fall of the Higgs field. Afterwards, 
the Higgs starts oscillating around our safe electroweak vacuum and promptly decays.
Meanwhile, Higgs quantum perturbations are generated. During inflation they leave the Hubble radius and the corresponding curvature perturbations freeze in, while the total curvature perturbation grows \cite{us}. 
The Higgs contributes to the curvature perturbation with a  peak at small scales, when there are about 20 e-folds to go till the end of inflation. 
During the radiation phase that immediately follows the end of inflation, the Higgs decays communicating its perturbations to the curvature perturbation now in the form of radiation. 
The final curvature perturbation is therefore flat on large scales, but has a peak at small scales. 
When these small-scale wave-lengths re-enter the horizon, if large enough, the may originate PBHs with a mass roughly equal to the mass contained in that horizon volume and with a mass fraction
\be
\label{fr}
\beta_{\rm prim}\simeq \frac{\Delta_c}{\sqrt{2\pi}\sigma_\Delta} e^{-\Delta_c^2/2\sigma^2_\Delta},
\ee
where
\be
\Delta(\vec x)=\frac{4}{9 a^2 H^2}\nabla^2\zeta(\vec x),
\ee
is the density contrast (during the radiation era), $a$ is the scale factor, $\Delta_c$ is the critical value above which a given region collapses to a PBH (its value is typically $\sim 0.45$), $\zeta$ is the gauge-invariant curvature perturbation, 
and we have defined the variance of the density contrast to be
\be
\sigma_\Delta^2(M)=\int{\rm d}\ln k \, W^2(k,R_H){\cal P}_\Delta.
\ee
Here $W^2(k,R_H)$  is a Gaussian window function smoothing out the density contrast on the comoving horizon length $R_H=1/aH$.
The present dark matter abundance made of PBHs of mass $M$ is therefore given by
\be
\left(\frac{\Omega_{\rm DM}(M)}{0.12}\right)\simeq\left(\frac{\beta_{\rm prim}(M)}{7\cdot 10^{-9}}\right)\left(\frac{106.75}{g_*}\right)^{1/4}\left(\frac{M_\odot}{M}\right)^{1/2},
\ee
where  $g_*$ is the effective number of degrees of freedom at the time of PBH formation.
In this expression we have neglected accretion which would eventually allow to start from smaller PBH mass fractions at formation. The results of Ref. \cite{us} have been repeated, confirmed and reported in Ref. \cite{strumia}.

\paragraph{The fine-tuning.}
As any other model of inflation which creates PBHs out of spiked perturbations at small scales, the scenario we have described is fine-tuned \cite{us,strumia}.
Indeed, the mass fraction (\ref{fr}) is exponentially sensitive to the variance of the density contrast. Small variations of it lead to too small or too large PBH abundances. In Ref. \cite{us} the fine-tuned choice of the parameters has been motivated anthropically. Structures can form through the dark matter under the form of PBHs and life can develop only in those  regions which survive the AdS catastrophe and are saved by the thermal effects. In this sense, the electroweak SM instability is a bonus.

Let us elaborate on fine-tuning. The abundance of  PBHs is sensitive to the initial condition of the Higgs field $h_*$  at the time when classicality takes over, that is   $\sim  20$ e-folds before the end of inflation.
Small deviations from $h_*$, $\delta h_* \simeq (10^{-3}-10^{-2})H$  lead to a too tiny value of the PBH abundance or to a fall into anti de Sitter. 
This variation has to be compared to the quantum fluctuations $\pm (H/2\pi)\simeq  0.16\,H$, which  arise on length scales $\sim H^{-1}$ during the first $e$-folds of evolution of the Higgs field.
This gives a small probability $\sim 10^{-2}$ for each Hubble volume  that the initial condition stays close to $h_*$. 

Since our observed universe (corresponding to a number of e-folds of about 60) contained at that time  about    $\exp(120)$ Hubble volumes, one might naively think that the total probability will be therefore $\sim  10^{-2 \exp(120)}$. This is not correct
as one is not interested in the probability of simultaneous production of PBHs in all Hubble size domains\footnote{In this sense, the production of PBHs as dark matter through our mechanism suffers  from a fine-tuning of the order of $\delta h_*/(\sqrt{2\pi}(H/2\pi))=\sqrt{2\pi}(\delta h_*/H)\simeq (10^{-3}-10^{-2})$ obtained assuming a Gaussian distribution for the Higgs field. This fine-tuning is typical of all mechanisms giving rise to PBHs through inflation.}.
The probability though applies to the counting of regions with might end up not being saved by the thermal effects. 
If one of those $\sim \exp(120)$ regions is not saved, it will expand after inflation and eventually engulf our entire universe. 
For each of these regions,  it was argued that the probability for the Higgs to be saved is of order $1/2$  \cite{strumia}.
The reason is that, in the regions that give an abundance of PBHs of the order of $\Omega_\text{DM}$, the Higgs field reaches a final value $\hc(\te)$ very close to the critical value not to create a dangerous AdS bubble. 
A small overfluctuation of $h_*$ or its initial velocity $\dot h(t_*)$ would push then $\hc(\te)$ into the AdS regime.
Thus, in Ref.~\cite{strumia} it is argued that the probability that none of the $\sim \exp(120)$ regions makes an AdS bubble is $2^{-\exp(120)}$.

In the published PRL version of our paper, now v2 of Ref. \cite{us}, it was already explicitly stated that the choice of  parameters  needed for PBH formation, although  fine-tuned,  would be motivated anthropically. 
The relevant issue is then the following: is $\sim  2^{-\exp(120)}$ really a small number from the point of view of the multiverse and anthropic argument? 
In fact, once one  accepts the anthropic principle, the  reasonable question is what one should multiply the tiny probability for? 
Within the eternal inflation/multiverse, one should use the volume-weighted physical probability which, unlike the comoving probability distribution, takes into account the overall growth of the volume of the universe: inflationary growth rewards parts of the universe with respect to others.
If one assumes a comoving probability point of view, a sample is assumed to be typical and then  general properties are deduced. 
However, distributions looking  atypical from an analysis based on the comoving probability, can be common when using the physical probability (and vice versa). 
We might well  live in a region of the  global universe which looks unusual if judged so using the comoving probability \cite{linde}. 
In other words, there might be a number of universes much bigger than $\sim  2^{\exp(120)}$  to probe.

As an example, we can refer to Ref. \cite{lindemultiverse} where it is estimated that the number of universes in eternal inflation is proportional to the exponent of the entropy of inflationary perturbations, $\exp(\exp(3N))$, where $N$  is the number of e-folds of slow-roll post-eternal inflation. 
If we assume  that our observed universe originates from only 60 e-folds of exponential expansion, one gets \cite{lindemultiverse}
\be
{\cal N}=\textrm{number of universes}\sim 10^{10^{77}}.
\ee
This is incomparably larger than the number ${\cal N}$ of universes needed to find, with a probability of order one, $n=\exp(120)$ adjacent regions which have not fallen into AdS. 
This probability can be approximated as $2^{-n}{\cal N}$, which gives ${\cal N}\gtrsim 2^{\exp(120)}\sim 10^{4\cdot 10^{51}}$. 
In chaotic inflation where the number of e-folds is typically $10^{12}$ one gets  \cite{lindemultiverse}
\be
{\cal N}\sim 10^{10^{10^{7}}}.
\ee
As scary as it might seem, the small probability quoted in Ref. \cite{strumia} takes an (exponentially) enormous advantage of this number of universes and  what seems unnatural in fact might turn out to  be natural.  For instance, as shown in Ref. \cite{tetradis}, bubbles of AdS may shrink during inflation, if they start with small radius and low velocity. Suppose that half of them shrink and half of them expand. This implies again that one just needs to live in that patch among the exponentially large number where all of the dangerous bubbles shrink and this probability in the multiverse will not be exponentially small, but order unity.

Furthermore, if one wishes to estimate the probability of survival of our universe, 
 it should also  be  remembered that luckily we live again in a period when the cosmological constant dominates. 
It can be easily calculated that the particle horizon in our universe from now until infinity will expand by just one third with respect to its current value. From that moment on, our universe will be screened
against AdS bubbles.

Leaving aside these considerations  which might render the reader (and us) uncomfortable for the  lack of any firm quantitative arguments, in this paper we will propose a natural solution to the fine-tuning problem. 
As mentioned above, the fine-tuning caused demanding the right abundance of PBHs is only $\sim (10^{-3}- 10^{-2})$. The problem arises when discussing the fine-tuning needed to save all the Hubble volumes when there are about 20 e-folds to go. 
So, one just needs a (reasonable) solution which will eliminate the presence of the AdS regions altogether without altering the attractive properties of the scenario, i.e. that the PBHs are generated by the SM Higgs and that dark matter is made of SM particles. This is what we will discuss in the next section.

\paragraph{Getting rid of the AdS regions altogether.}
As we stressed in Ref. \cite{us} the mechanism to produce PBHs, which today  form the dark matter without resorting to any dark matter particle beyond the Standard Model, relies on the  instability of the electroweak vacuum, so that the Higgs perturbations can grow   during inflation. This dynamics is totally built within the Standard Model.
On the other hand, to get rid of the fine-tuning caused by the dangerous AdS vacua, one can simply alter the form of the Higgs potential at   energies much larger than the instability scale, thus  without
altering the nice features of the mechanism. 

Let us then suppose that at energy scales  much larger than the instability scale $\sim 10^{11}$ GeV,  there are new particles whose interaction with the Higgs can  change the sign of the quartic coupling from negative to positive again, thus stabilising the Higgs  potential.
As a simple case, add a complex scalar field $\S$ with potential \cite{es}
\begin{equation}
V = \lambda_S (|S|^2 - \omega^2/2)^2 + 2\lambda_{HS}  (|S|^2 - \omega^2/2) (|\varphi_H|^2 - v^2/2)\, ,
\label{eq: Lagrangian}
\end{equation}
where $S$ is the additional singlet (with vacuum expectation $\omega$) and $\varphi_H$ the SM Higgs doublet (with vacuum expectation $v$). 
The interaction term in Eq.~\eqref{eq: Lagrangian}, generates a threshold contribution $\delta\lambda=-\lambda_{HS}^2/\lambda_S$ to the Higgs quartic $\lambda(h)$ above a scale $\sim \mS$ \cite{es}. Moderate values of $\lambda_{HS}$
and $\lambda_S$ can produce a $\delta\lambda$  large enough to bring $\lambda(h)$ to be positive for $h\gtrsim \mS=\sqrt{\lambda_S}\omega$.  In such case the Higgs potential displays a true minimum at $h\sim \mS$.

For our purposes we consider the case $\mS\sim \mathcal O(\TRH)$.
In this way, even if the Higgs jumps beyond the barrier early during inflation, it will stop at its true minimum and thermal effects at reheating rescue the Higgs, bringing it back towards the electroweak vacuum.

\begin{figure}[t!]
\includegraphics[width=1.\columnwidth]{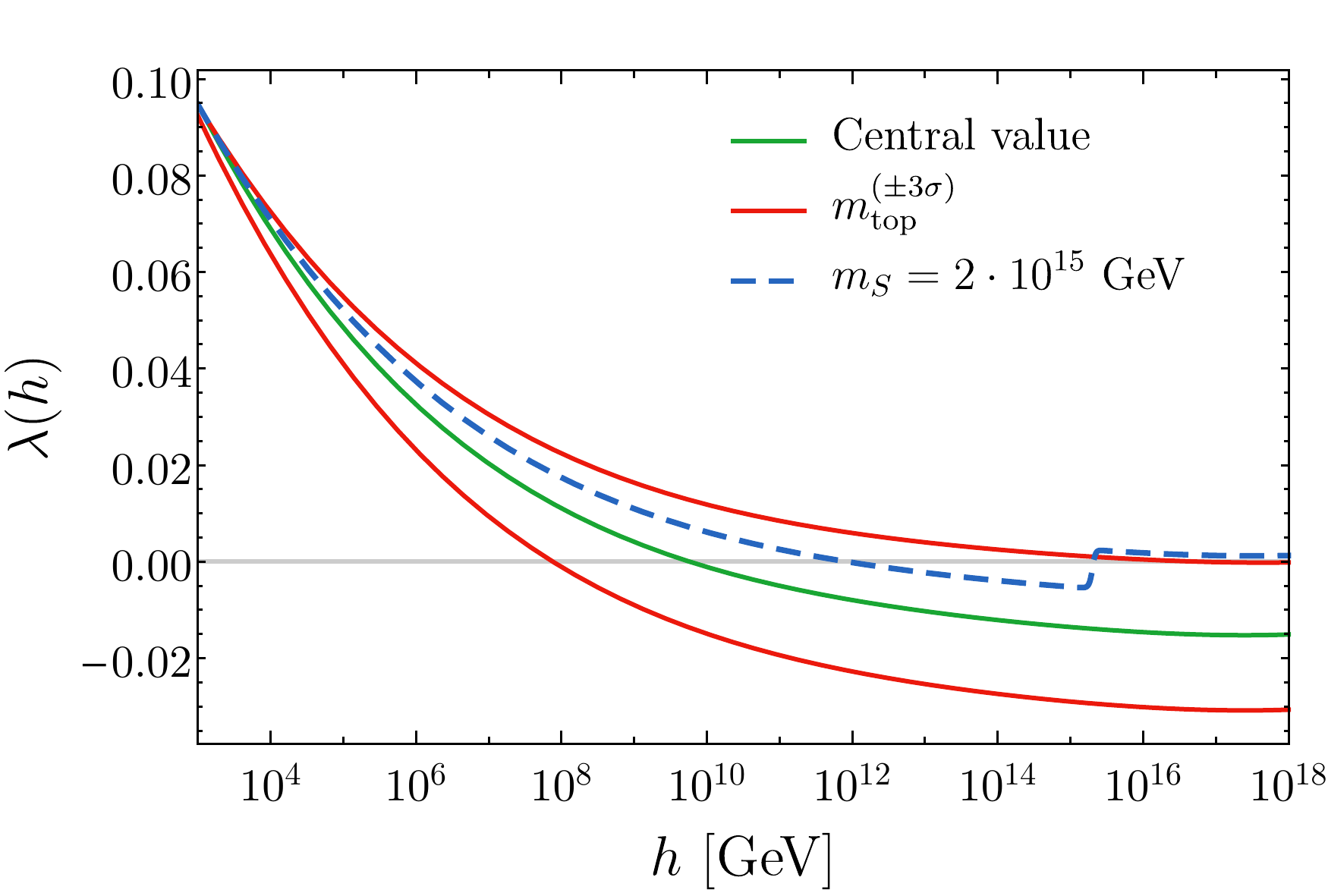}
\caption{Running of the Higgs quartic coupling $\lambda$ with the introduction of an extra scalar (dashed blue line), together with the central and marginal values within the SM \cite{instab}.}
\label{fig: running lambda}
\end{figure}

We show in Fig.~\ref{fig: running lambda} the running of $\lambda$ corresponding to $\mS=2\cdot 10^{15}$ GeV, and $\lambda_{SH} = 0.05$, $\lambda_S=0.3$.
We stress that there is no fine-tuning here: we can allow relative variations of order $(10-20)\%$ for $\mS$, and the only requirement for $\lambda_{SH}$ and $\lambda_S$ is that they yield $\lambda(h\gtrsim\mS)>0$, corresponding to $|\delta\lambda| \gtrsim 0.008 $.
\newline
We repeat  the analysis in Ref. \cite{us}, with the same choice of the Hubble rate $H=10^{12}$ GeV and $m_\textrm{top}= 172$ GeV.
The Higgs field starts its classical evolution beyond the barrier  at $10^{11}$ GeV from a value $h_*$ at the time $t_*$ (corresponding to $N_*$ $e$-folds till  the end of inflation).
We solve the equations of motion for the Higgs background $\hc$ and its perturbations $\delta h_k$, and compute the power spectrum $\Pz$ of the comoving curvature perturbation $\zeta$.

\begin{figure}[t!]
\includegraphics[width=1.\columnwidth]{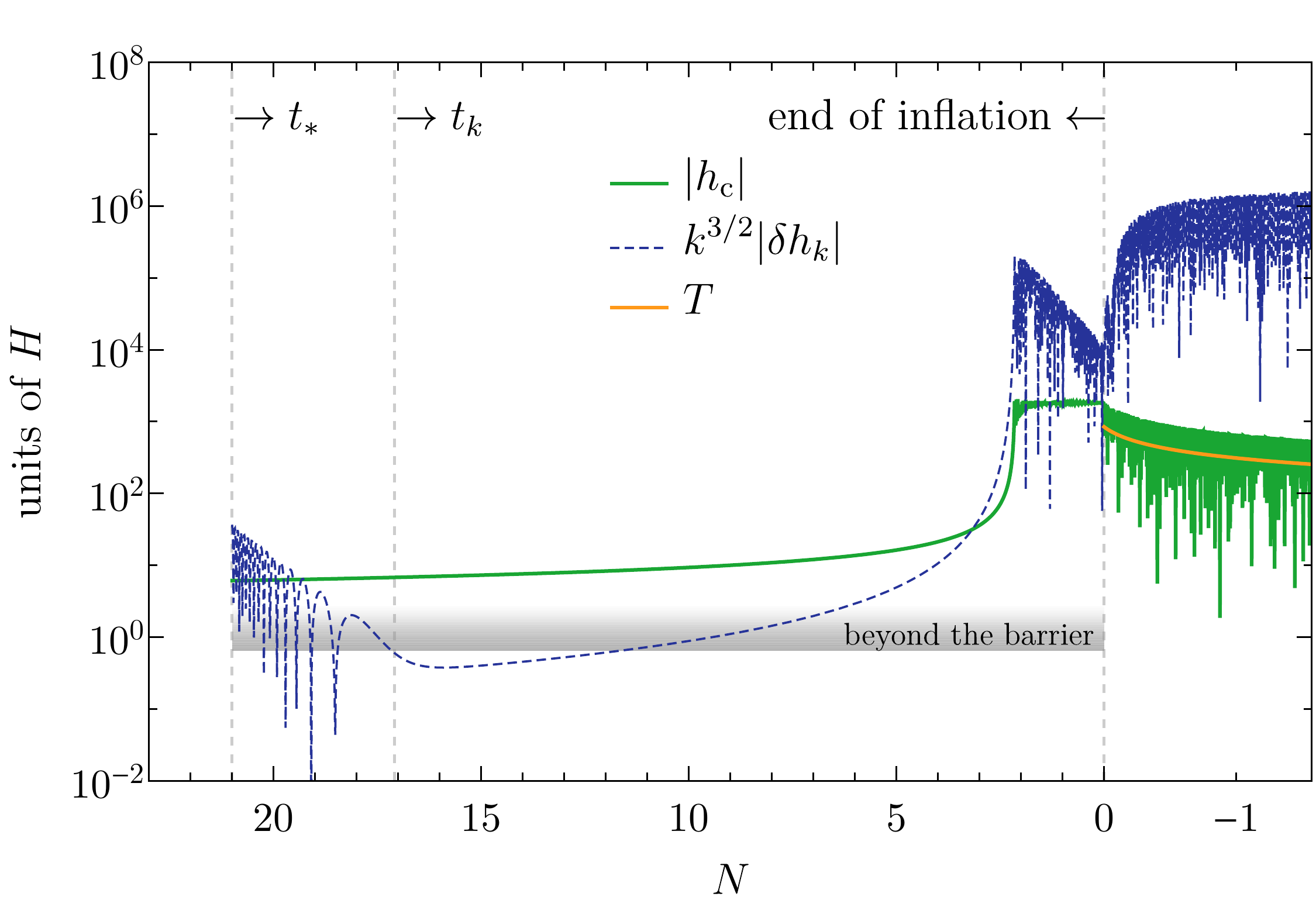}
\caption{Evolution of $\hc$ and its perturbation $\delta h_k$ for the case $N_*>\Nstar$ (or equivalently $h_*>\hstar$; see the text for details).}
\label{fig: evolution}
\end{figure}

The outcome is the following. 
Let us fix for the moment $h_*$, and denote by $\Nstar$ the initial time which would give the right abundance of PBHs without the presence of $\S$.
By including $\S$, the Higgs potential does not change for $h<\mS$, so that for $N_*\leq \Nstar$ the evolution of the Higgs is not altered with respect to what was discussed in Ref. \cite{us}: for $N_*<\Nstar$, the final $\Pz$ is too small to seed PBH.
For $N_*>\Nstar$, the Higgs field reaches its minimum at $\mS$ before the end of inflation, and starts oscillating around it\footnote{The Higgs mass in this true minimum is typically larger than $H$.}. 
In the meantime, the tachyonic excitation of the Higgs fluctuations ceases, and $\delta h_k$ oscillates around zero with the same frequency as $\hc$. 
The evolution for this case $N_*>\Nstar$ is shown in Fig.~\ref{fig: evolution}.

\begin{figure}[t!]
\includegraphics[width=1.\columnwidth]{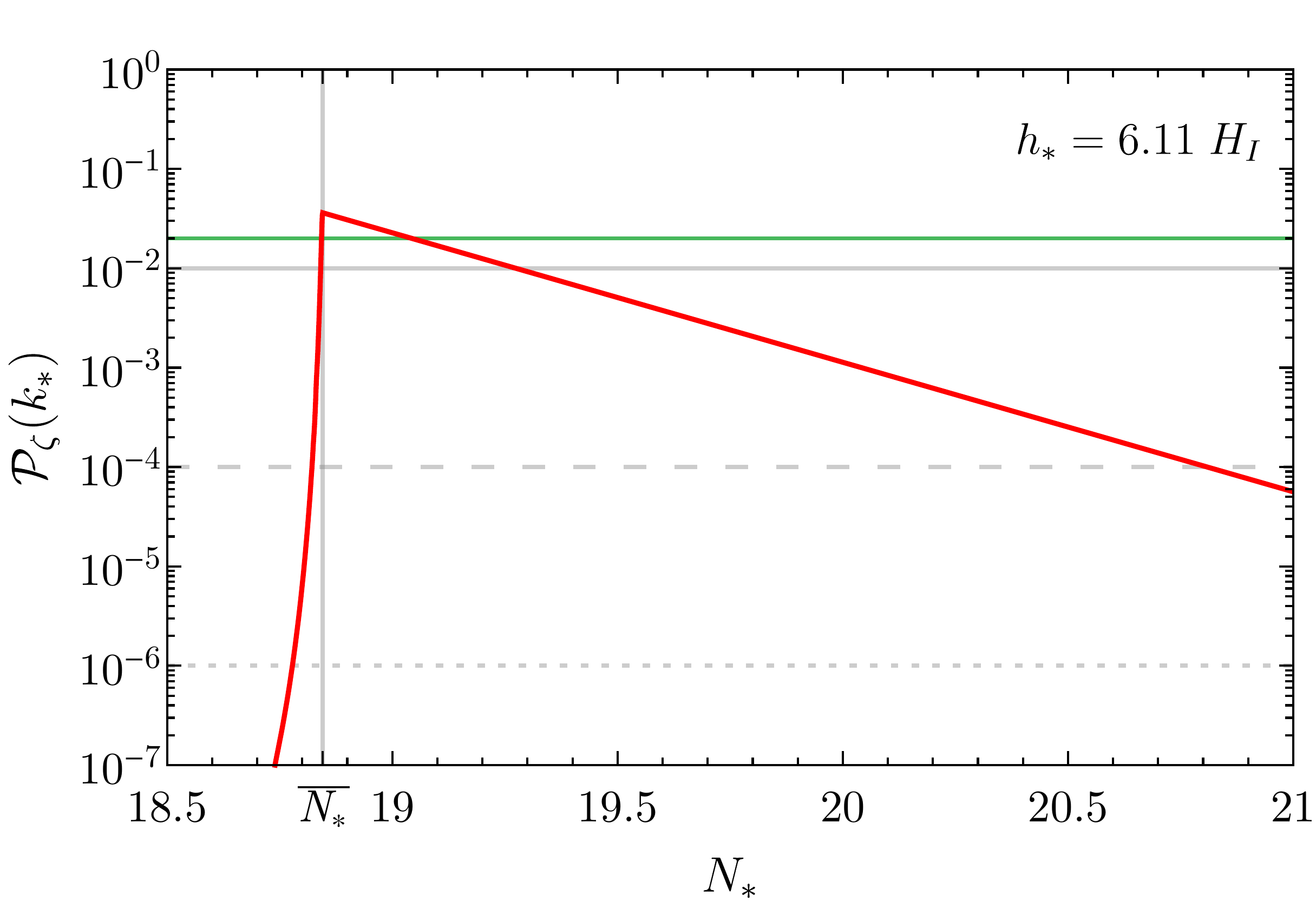}
\caption{Power spectrum $\Pz$ as a function of the starting value $N_*$, for a fixed $h_*=6.11\,H$.
The green line corresponds to a PBH abundance roughly of order unity, and the grey lines, yielding no PBHs, correspond to the cases shown in Fig.~\ref{fig: fine tuning}.}
\label{fig: Nstar Pz}
\end{figure}

\begin{figure}[t!]
\includegraphics[width=1.\columnwidth]{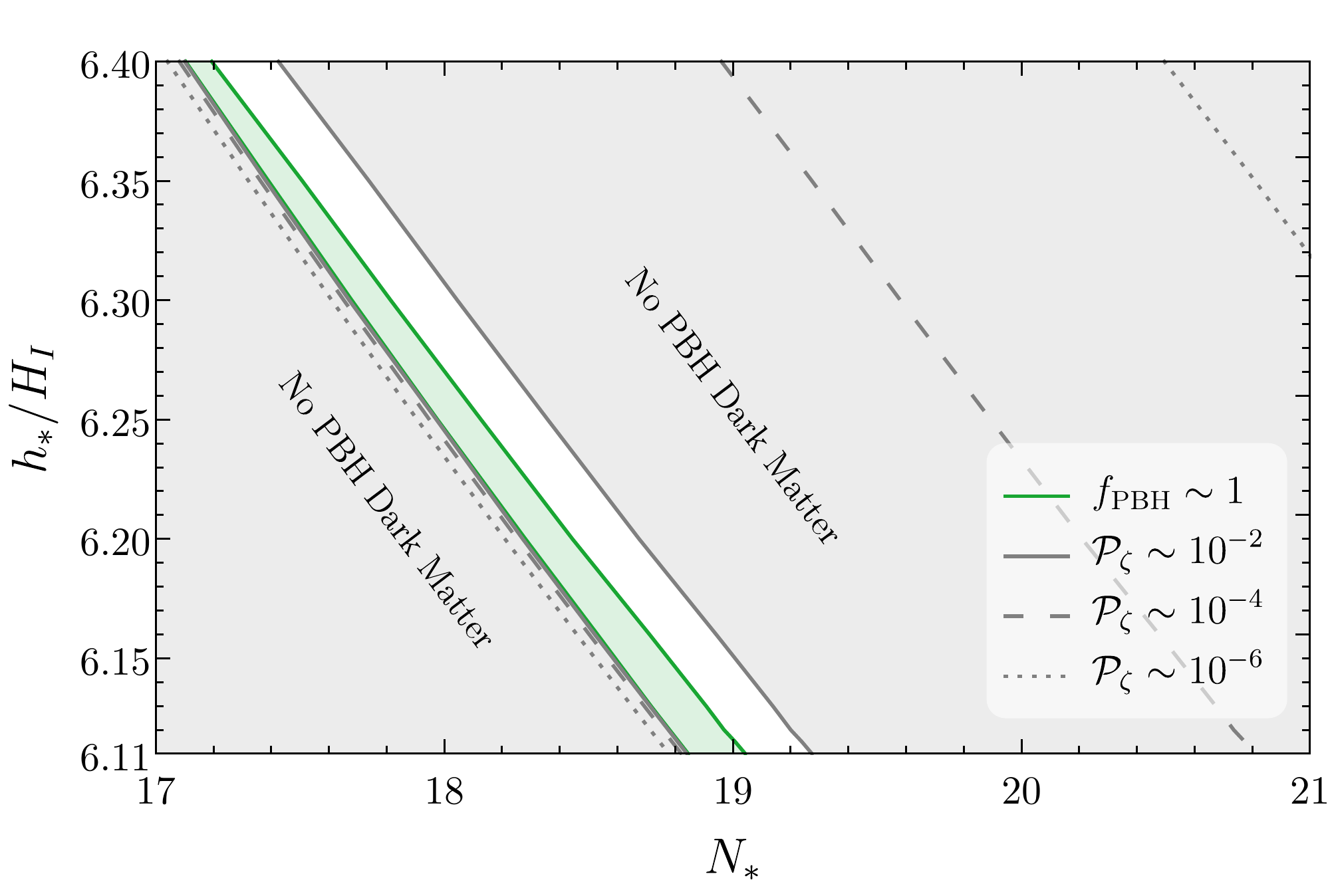}
\caption{Power spectrum $\Pz$ as a function of the starting values $N_*$, $h_*$. 
The grey regions do not yield PBHs, and the green region corresponds to a non-negligible PBH abundance.}
\label{fig: fine tuning}
\end{figure}

The redshift of $\delta h_k$ during this phase (in which the Higgs behaves as a matter fluid) slowly reduces the amplitude of $\zeta$ as $a^{-3/2}$, and $\Pz$ decreases as $a^{-3}$ during the oscillation of $\hc$ around its true minimum.\footnote{Both in Ref. \cite{us} and in this paper we assume for simplicity a constant Hubble rate during inflation, in order not to specify an inflation model. 
This implies $\dot \rho_\textrm{tot}=\dot \rho_h$ during inflation. This approximation works less well during the oscillating phase in which the Higgs behaves as matter, and $\rho_h\sim a^{-3}$. 
In any case the qualitative behaviour of $\Pz$ shown in Fig.~\ref{fig: Nstar Pz} and \ref{fig: fine tuning} would be the same: the slope of $\Pz$ for $N_*>\Nstar$ in Fig.~\ref{fig: Nstar Pz} would be steeper, without altering our conclusions.}
The final value of $\Pz$ for the mode $k_*=a(t_*)H$ (which exits the Hubble radius at $N_*$) is shown in Fig.~\ref{fig: Nstar Pz}.\\

Notice that the previous discussion proceeds in the same way if we  fix a generic $N_*$ and identify what $\hstar$ leads to the right PBH abundance without the presence of $\S$. 
For $h_*\leq\hstar$ the evolution is the same described in Ref. \cite{us}, whereas for $h_*>\hstar$ the curvature perturbation is slowly reduced at the end of inflation.

Fig.~\ref{fig: fine tuning} shows the final power spectrum $\Pz$ as a function of the initial values $N_*$ and $h_*$.
On the left-hand side of the plot we have the same situation as in Ref.~\cite{us}, with a strong variation of the final $\Pz$ even for per mille variations of $N_*$ and $h_*$.
On the right-hand side of the plot we find the region which would have fallen into AdS without the threshold correction on $\lambda$ at $\mS$.
If $\lambda$ is pushed to positive values at a scale close to $\TRH$, then the Higgs is always rescued and there are no AdS regions which could form.
Moreover, the dependence of $\Pz$ on $N_*$ is much milder on the right side of Fig.~\ref{fig: fine tuning}, and a second region leading to the right PBH abundance is found.
We highlight all the region corresponding to $\Pz(k_*)\sim (0.02-0.04)$: although values larger than $\sim 0.02$ yield a too large PBH abundance, we  recall the argument exposed previously. The formation rate of PBH is a probabilistic quantity and says nothing about the distribution of such BH in space.
Even if some patches lead to a larger abundance they could be compensated by regions without PBH. 
These inhomogeneities occur at very small scales and do not constitute a  problem as the PBHs are generated anyway strongly clustered \cite{c}.

\bigskip
\paragraph{Conclusions.}
In this short note we have addressed the issue of fine-tuning affecting the mechanism proposed in Ref. \cite{us}. Be the  fine-tuning explained or not 
within the multiverse paradigm, we have discussed a natural and economic alternative to get rid of it. We are convinced that the 
cornerstones the scenario is built on, above all the fact that it is built on the physics of the SM and it does not require physics beyond the SM to explain dark matter, remain robust.

\bigskip
\paragraph{Acknowledgments.}
We thank A. Linde  for discussions. A.R. and D.R. are supported by the Swiss National Science Foundation (SNSF), project {\sl Investigating the Nature of Dark Matter}, project number: 200020-159223.
The work of J.R.E. has been partly supported by the ERC
grant 669668 -- NEO-NAT -- ERC-AdG-2014, the Spanish Ministry MINECO under grants  2016-78022-P and
FPA2014-55613-P, the Severo Ochoa excellence program of MINECO (grants SEV-2016-0588 and SEV-2016-0597) and by the Generalitat de Catalunya grant 2014-SGR-1450.

\end{document}